# Transport evidence of robust topological surface state in BiTeCl single crystals, the first strong inversion asymmetric topological insulator

F.X. Xiang, X.L. Wang*, and S.X. Dou

*Spintronic and Electronic Materials Group, Institute for Superconducting and Electronic Materials, Australian Institute for Innovative Materials, University of Wollongong, North Wollongong, 2500, Australia*

**Three-dimensional (3D) topological insulators (TIs) are new forms of quantum matter that are characterized by their insulating bulk state and exotic metallic surface state, which hosts helical Dirac fermions[1,2]. Very recently, BiTeCl, one of the polar semiconductors, has been discovered by angle-resolved photoemission spectroscopy to be the first strong inversion asymmetric topological insulator (SIATI). In contrast to the previously discovered 3D TIs with inversion symmetry, the SIATI are expected to exhibit novel topological phenomena, including crystalline-surface-dependent topological surface states, intrinsic topological *p-n* junctions, and the pyroelectric and topological magneto-electric effects[3]. Here, we report the first transport evidence for the robust topological surface state in the SIATI BiTeCl via observation of Shubnikov-de Haas (SdH) oscillations, which exhibit the 2D nature of the Fermi surface and $\pi$ Berry phase. The $n = 1$ Landau quantization of the topological surface state is observed at $B \approx 12$ T without gating, and the Fermi level is only 58.8 meV above the Dirac point, which gives rise to small effective mass, $0.055m_e$, and quite large mobility, 4490 $cm^2s^{-1}$. Our findings will pave the way for future transport exploration of other new topological phenomena and potential applications for strong inversion asymmetric topological insulators.**

The discovery of topological insulators (TIs) has aroused great interest in the condensed matter physics community because of their exotic conducting surface state, with properties such as helical spin texture[4,5], immunity to backscattering[6], giant and linear magnetoresistance (MR)[7], zeroth order Landau quantization[8,9], etc., which originate from the strong spin-orbit coupling and are protected by time-reversal symmetry[1,2]. Breaking of the time-reversal symmetry can result in the realization of Majorana fermions[10] and the anomalous quantum Hall effect[11,12]. Inversion symmetry plays an important role in searching for topological insulator candidates by the parity criteria, in such cases as $BiSb_{1-x}$, $Bi_2Se_3$, and $Bi_2Te_3$[13-16]. In contrast, inversion asymmetric topological insulators (IATIs) are theoretically predicted to exhibit new and unique topological phenomena, including topological *p-n* junctions, surface states at certain crystalline surfaces, electric polarization or net charges, pyroelectricity, and the magneto-electric effect, as well as topological superconductivity[17]. It is highly desirable to find strong IATIs with large net charges. The only candidate so far has been strained HgTe, in which the inversion symmetry is broken by its zinc blende crystal structure, but no net charge polarity is present[18]. Strained HgTe is therefore called a weak inversion asymmetric topological insulator (IATI). Very recently, BiTeCl was discovered, by an angle resolved photoemission spectroscopy (ARPES) study, to be the first strong inversion asymmetric topological insulator, in which positive and negative charge occurs at Cl and Te termination, respectively[3,19]. For the samples with Te termination, the Dirac point of the topological surface state is located 220 meV below the edge of the bulk conduction band (Fig. 1). Because its inversion symmetry is naturally broken by its crystal structure, it is only necessary to break the time-reversal symmetry to realize the topological magneto-electric effects[13,18]. Its large bulk gap also holds promise that the new topological phenomena can be realized at room temperature[3].

There is an urgent need to see transport confirmation of the topological surface state, which is paramount before further exploration for other transport topological electronic states can be undertaken. Here, we report the first transport evidence for the topological surface state in the strong IATI (SIATI) BiTeCl via observation of Shubnikov-de Haas (SdH) oscillations, which exhibit the two-dimensional (2D) nature of the Fermi surface and the π Berry phase.

Quantum oscillations have already been employed to detect and analyze the topological surface state and other non-trivial Berry phase systems[20-28]. In the scenario of semiclassical physics, the presence of an external magnetic field causes the cyclotron motion of electrons, which are quantized into various discrete energy levels, termed Landau levels. By varying the magnetic field or electron density, the Landau levels shift and finally cross the Fermi surface, resulting in quantum oscillation. The extremal cross-section of the Fermi surface and the Landau level index *n* are correlated by

$$2\pi(n+ \beta +1/2)=S_F\hbar/eB, \qquad (1)$$

where *n* is the Landau level index, $S_F$ is the extremal cross-section of the Fermi surface, and *β* is the phase factor, which is crucial for identifying the topological surface state. *β* = 0 corresponds to the trivial Berry phase of free fermions, and *β* = 1/2 corresponds to the π Berry phase of Dirac fermions. Such a non-trivial Berry phase can be accessed by studying the transport manifestation of quantum oscillations, SdH oscillations. According to the Lifshitz-Kosevich formula,

$$R_{xx} \propto R_T R_D \cos 2\pi(\frac{F}{B} + \beta + \frac{1}{2}) \qquad (2)$$

where *F* is the frequency of SdH oscillations as a function of 1/*B*, which equals $S_F\hbar/eB$, $\omega_c$ is the cyclotron frequency, and $\varphi_B$ is the Berry phase, $R_T = \alpha T m^*/B sinh(\alpha T m^*/B)$ is the thermal damping factor, and $R_D = exp(-\alpha T_D m^*/B)$ is the Dingle damping factor, with *sinh* the hyperbolic function, the effective mass $m = m^* m_e$, $\alpha = 2\pi^2 k_B/e\hbar \approx 14.69$ T/K, and $T_D = \hbar/2\pi k_B \tau$ is the Dingle temperature, with *τ* the scattering time.

Here, we report the experimental investigation of magneto-transport in BiTeCl single crystals, the first strong inversion asymmetric topological insulator. By applying magnetic fields up to 13.5 T, we observed the SdH oscillations, which indicate the 2D nature of the Fermi surface, and the $\pi$ Berry phase, which confirms the presence of the topological surface state in the BiTeCl. Furthermore, when the Fermi level is close to the Dirac point, the topological surface electrons exhibit a small effective mass, $0.055m_e$, and high carrier mobility, 4490 cm$^2$s$^{-1}$. Our transport results verify the topological surface state discovered by the ARPES study[3].

The as-grown single crystals are plate-like (inset in Fig. 1(a)), have dimensions up to 5 × 10 mm$^2$ with different thicknesses, and can be easily cleaved mechanically. Fig. 1(a) presents the X-ray diffraction pattern of a typical BiTeCl single crystal. Only (00*l*) peaks are present, which means that the *c*-axis is perpendicular to the cleaved plane. The full width at half maximum (FWHM) of the (002) peak is as small as 0.087°, indicating the high quality of the single crystals. Fig.1b shows the schematic band structure in our three samples, the Fermi surfaces of which are indicated by the blue solid lines in terms of the standard SdH oscillation analysis as carried out below.

To detect the topological surface state via the SdH oscillations, external magnetic field was applied up to 13.5 T, and clear SdH oscillations could be detected from the 1$^{st}$ derivative of the MR at 10 K for S1 and 2.5 K for S2 and S3. Firstly the dimensional feature of the Fermi surface is analysed by measuring the SdH oscillation at various tilt angles θ, which are defined as the angles between the magnetic field and *c* axis of crystal with the magnetic field always perpendicular to electric current direction. Because the Fermi surface of topological surface state is two-dimensional (2D), the periodicity of the oscillation only depends on $1/B\perp$, where $B\perp = B\cos\theta$ is the perpendicular component of magnetic field. Fig. 2a-c, show the dR/dB versus $1/B\perp$ for S1, S2 and S3, respectively. It can be seen that the peaks of the 1$^{st}$ derivative marked by the dashed lines only depend on $1/B\perp$, which indicates that the SdH oscillations originate from a 2D system. Secondly, to demonstrate that whether or not the 2D system is the topological surface state, the Landau level fan diagram is plotted. Because topological surface state is protected by the time-reversal symmetry, the electrons acquire a $\pi$ Berry phase after circling the Fermi surface[1], which can be determined via the analysis of the Landau level fan diagram. As shown in Fig. 2d the Landau level is plotted as a function of 1/B, where the minimum of the SdH oscillation is located at the Landau level index *n* which corresponds to the Hall plateau in the quantum Hall effect limit, while the maximum is located at the Landau level index *n*+1/2. As can be seen in the inset of Fig. 2(d), the linear fits to the data yield an intercept with the *n*-index axis at -1/2, which corresponds to the $\pi$ Berry phase, as described in Equations (1) and (2). Therefore, we conclude that the SdH oscillations originate from the topological surface state in our SIATI BiTeCl single crystals.

After verification of the topological surface state in BiTeCl, further analysis was carried out to gain insight into the transport information on the topological surface state in the BiTeCl compound. According to Equation (1), the slope of the linear fitting lines in Fig. 2 yields the oscillation frequency *F* = 14.04, 121, and 156 T for samples S1, S2, and S3, respectively, which are consistent with oscillation frequencies obtained from the fast Fourier transform. Because of the circular Fermi surface near the Dirac point of BiTeCl, the extremal cross-section of the Fermi surface can be written as $S_F = \pi k_F^2$; based on the oscillation frequency $F = (\hbar c/2\pi e)S_F$, the Fermi surface radius, $k_F$ = 2.06 × 10$^8$, 6.06 × 10$^8$, and 6.88 × 10$^8$ m$^{-1}$ is obtained for S1, S2, and S3, respectively. For the 2D electron system, the carrier density $n_s = k_F^2/4\pi$, so $n_s$ of the topological surface state for S1, S2 and S3 is 0.34 × 10$^{12}$, 2.92 × 10$^{12}$, 3.77 × 10$^{12}$ cm$^{-2}$, respectively.

For the quantum oscillations, the temperature ($k_BT$) and magnetic field ($\hbar\omega$) can affect the resolution of the Landau tubes, as lower temperature and higher magnetic field can lead to larger oscillation amplitude. This is well described by the Lifsitz-Kosevich theory with the thermal damping factor $R_T$ and the Dingle damping factor $R_D$, as described in Equation (2). Therefore, the SdH oscillations were measured at various temperatures. Clear SdH oscillations for S1, S2, and S3 were obtained after subtracting the smooth background, as shown in Fig. 3(a-c). It can be seen that the oscillation amplitude decreases as the temperature increases, because the thermal vibration blurs the Landau quantization. Fig. 3(d-f) shows the temperature dependence of the normalized oscillation amplitude for S1, S2, and S3, respectively. The amplitudes are obtained around 11.1, 13.3, and 12.5 T for S1, S2, and S3, and normalized by the extrapolated amplitudes at 0 K, respectively. The solid lines are the fitting lines based on the thermal damping factor in Equation (2), which yields the cyclotron effective mass $m = 0.055m_e$, $0.135m_e$, and $0.13m_e$ for S1, S2, and S3, respectively. Then, the Fermi velocity can be calculated as $v_F = \hbar k_F/m = 4.34 \times 10^5$, $5.40 \times 10^5$, and $6.13 \times 10^5$ m s$^{-1}$, for S1, S2, and S3, respectively, which is consistent with the ARPES data[3].

According to the ARPES investigation, the charge carriers could be *n*-type or *p*-type for samples with Te and Cl termination, respectively. The Hall measurements, however, indicate that the charge carriers are *n*-type in all of our measured samples. Using $E_F = mv_F^2$, where $E_F$ is the Fermi energy, the Fermi levels of S1, S2, and S3 are determined to be 58.8, 223.8, and 277.8 meV above the Dirac point. Figure 1(b) shows a schematic diagram of the band dispersion for S1, S2, and S3, with the Fermi levels indicated by the solid lines. It should be noted that the Fermi level of S1 is very close to the Dirac point.

Because $\Delta R/R_0 \approx R_D R_T$, while $\Delta RB\sinh(\alpha Tm^*/B) = \alpha Tm^*/\exp(-\alpha T_D m^*/B)$, by plotting the semi-log plot of $\Delta RB\sinh(\alpha Tm^*/B)$ versus $1/B$, a linear relationship is expected to be obtained, as shown in Fig. 3(g-i). The solid lines are the linear fitting lines, all with the same slope, which yields the Dingle temperature $T_D = 6.81$, 53.94, and 47.13 K. So, the scattering times are $\tau = \hbar/2\pi k_B T_D = 1.79 \times 10^{-13}$, $0.227 \times 10^{-13}$, $0.258 \times 10^{-13}$ s, the mean free paths are $l_s = v_F \tau = 5.48 \times 10^{-8}$, $1.58 \times 10^{-8}$, and $1.21 \times 10^{-8}$ m, and the carrier mobilities are $\mu = el_s/\hbar k_F = 4490.0$, 348.7, and 303.2 cm$^2$s$^{-1}$ for S1, S2, and S3, respectively. The transport parameters obtained from the SdH oscillation analysis are given in Table 1.

Our SdH oscillation analysis reveals that the transport parameters for S2 and S3 are similar to what has been reported for the inversion symmetry topological insulators. Sample S1, however, shows some unique features. First, around $B = 12$ T, the topological surface state is quantized into the $n = 1$ Landau level without gating as marked by the dash line in Fig.2d, which can only be observed in much higher magnetic fields (> 30 T) in inversion symmetric topological insulators such as $Bi_2Te_3$, $(Bi_{1-x}Sb_x)_2Se_3$ and $Bi_2Te_2Se_3$[20,29,30], implying that the Fermi level is very close to the Dirac point and low electron density. This would be an advantage for realizing the zeroth Landau level quantization of the topological surface state in the BiTeCl. Second, the electrons of the topological surface state in S1 exhibit superior transport performance with long scattering time, long mean free path, and high carrier mobility compared to the samples S2 and S3, due to the linear energy-momentum dispersion relationship near the Dirac point and the immunity to backscattering.

It should be emphasized that all of the transport measurements were performed under ambient atmosphere. We found that the topological surface state can be easily detected and is repeatable in many samples that we have measured, indicating that the topological surface state in BiTeCl compound is robust and inert to ambient atmosphere. This makes BiTeCl an excellent candidate for future transport exploration of other new topological phenomena and potential applications for this compound.

In summary, we have observed SdH oscillations in BiTeCl single crystal. The angular dependence of the magnetoresistance shows the 2D nature of the oscillation. Analysis of the Landau level fan yields the π Berry phase, providing transport evidence for a robust topological surface state in the strong inversion asymmetric topological insulator BiTeCl. $n = 1$ Landau quantization of the topological surface state is observed at $B \approx 12$ T. The topological surface state electrons close to the Dirac point exhibit superior transport performance with lower electron density, long scattering time, long mean free path, and high carrier mobility.

**Methods**

High quality single crystals of BiTeCl were grown by the self-flux method according to the $Bi_2Te_3$-$BiCl_3$ binary phase diagram in an evacuated quartz tube. First, $Bi_2Te_3$ was synthesized from high-purity (5N) Bi and Te powders. Due to the peritectic melting characteristic of BiTeCl, ground $Bi_2Te_3$ and $BiCl_3$ (5N) powders were weighed out with the molar ratio of 1:9 and thoroughly ground together, which was carried out in an oxygen and moisture monitored glove box to prevent the deliquescence of $BiCl_3$. The mixture of powders then was loaded into the quartz tube, which was sealed under vacuum. The sealed quartz tube then was heated above 420 °C over several hours, maintained there for 12 h, and then slowly cooled down to 200 °C over several days. The plate-like crystals were obtained by chemically removing the residual flux of $BiCl_3$.

Single crystal samples with shining surfaces were cleaved from the as-grown crystals and used for the standard four probe transport measurements. Gold wires were attached to the sample surface by silver epoxy, which was cured at room temperature before measurements to ensure the Ohmic contacts. Magnetic field was applied perpendicular to the sample surface up to 13.5 T. The angular dependence of the magnetoresistance was measured for determining the 2D nature of the Shubnikov-de Haas oscillations.

*Acknowledgements*

X.L.W. acknowledges the support from the Australian Research Council (ARC) through an ARC Discovery Project (DP130102956) and an ARC Professorial Future Fellowship project (FT130100778).

*Correspondence and requests for samples should be addressed to X.L.Wang (xiaolin@uow.edu.au).*

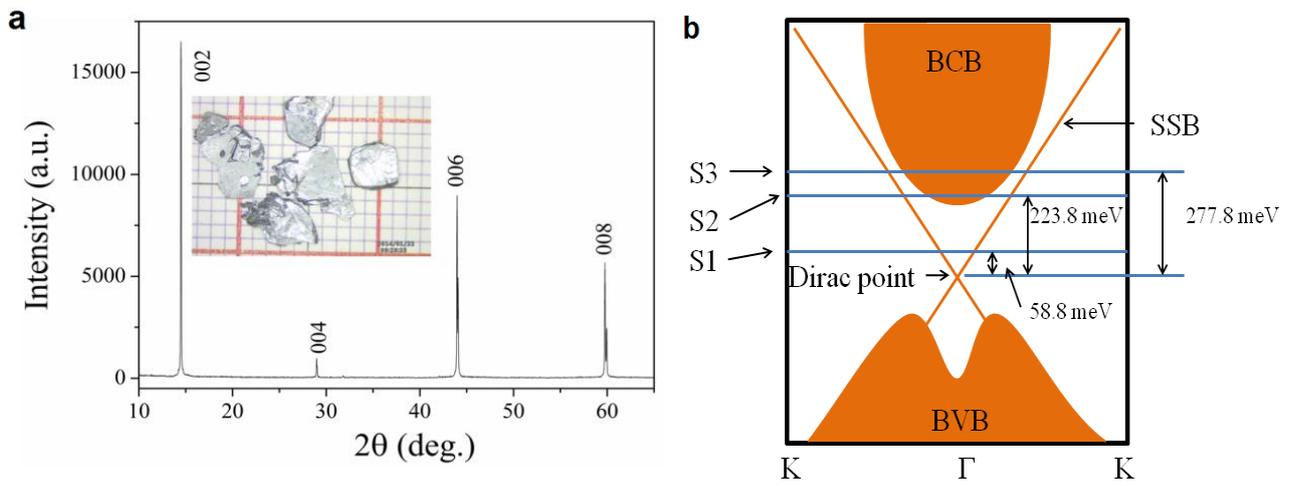

**Figure 1 | X-ray diffraction pattern of a typical single crystal and schematic band dispersion of BiTeCl: a**, XRD pattern of BiTeCl single crystal. Only (00l) peaks appear, indicating that the crystal surface is perpendicular to the *c* axis. The inset is a photograph of typical single crystals against a 1 mm scale. **b**, Schematic diagram of band dispersion [traced from reference (3)]. The Dirac point and Fermi levels of samples S1, S2, and S3 are marked by the blue lines.

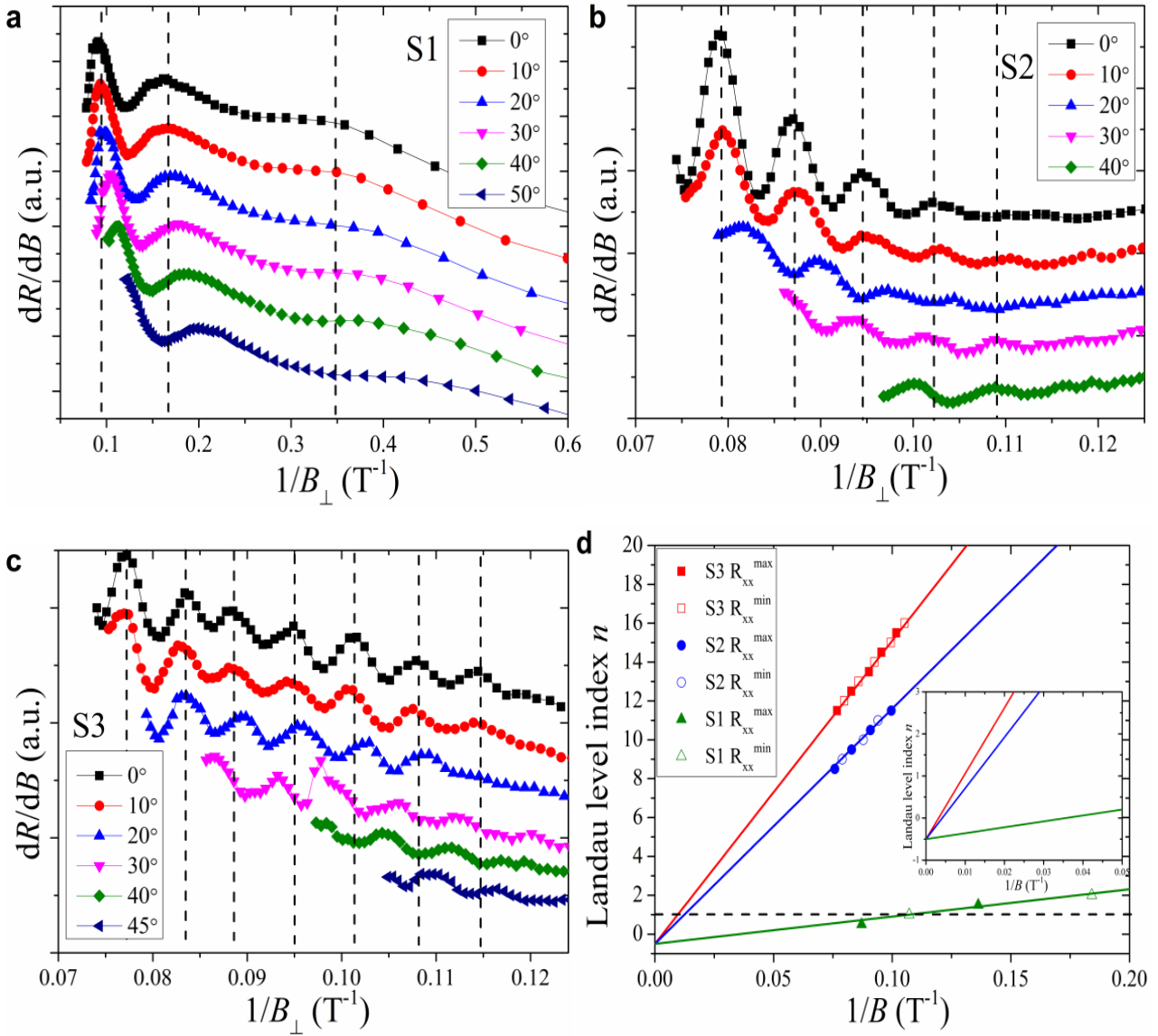

**Figure 2 │ Transport evidence of topological surface state in BiTeCl from the field dependence of the magnetoresistance (*R*): a-c,** Derivative d*R*/d*B* versus $B_\perp$ for S1(**a**), S2(**b**), and S3(**c**) at various tilt angles. The tilt angle *θ* is defined as the angle between the magnetic field and the *c*-axis, and the vertical component of the magnetic field $B_\perp = B\cos\theta$. The d*R*/d*B* peaks marked by the dashed lines depend only on $1/B_\perp$, which verifies the 2D nature of the Fermi surface. **d,** Landau level fan diagram for S1, S2, and S3. The solid lines are the fitting lines, the slopes of which are consistent with the oscillation frequencies obtained from the fast Fourier transform. The intercepts of the three fitting lines are -0.5 in terms of the Landau level index, as shown in the inset, which verifies the presence of the π Berry phase.

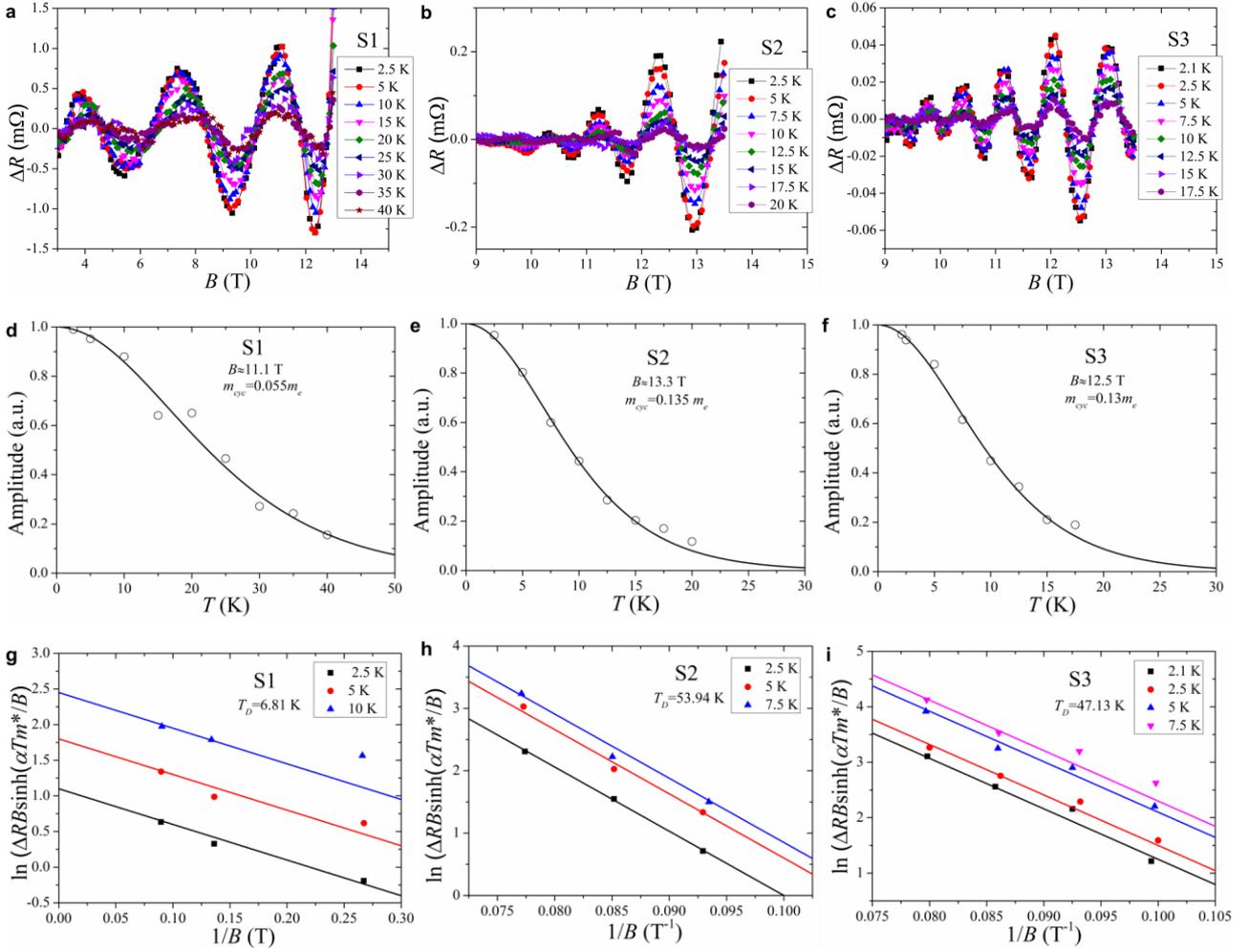

**Figure 3 | Extracting information of topological surface state via temperature dependent SdH oscillations: a-c**, SdH oscillations at various temperatures after subtracting the smooth magnetoresistance background. **d-f**, Temperature dependence of oscillation amplitude normalized by the extrapolated amplitude at 0 K. Fitting the data with the thermal damping factor in Equation (2) yields the effective masses, $0.055m_e$, $0.135m_e$, and $0.130m_e$, for S1, S2, and S3, respectively. **g-i**, Dingle plot of $\log[\Delta RB\sinh(\alpha Tm^*/B)]$ vs. $1/B$ at various temperatures, which shows the linear relationship. The slopes of the linear fitting lines give the Dingle temperature $T_D$ as 6.81 K, 53.94 K, and 47.13 K for S1, S2, and S3, respectively.

**Table 1 | Obtained parameters in the SdH oscillation study for S1, S2, and S3.**

|    | $F$ T | $k_F$ ×10⁸ m⁻¹ | $n_s$ ×10¹² cm⁻² | $v_F$ ×10⁵ ms⁻¹ | $m^*$ | $E_F$ meV | $T_D$ K | $\tau$ ×10⁻¹³ s | $l_s$ ×10⁻⁸ m | $\mu$ cm²V⁻¹s⁻¹ |
|----|------|------|------|------|------|------|------|------|------|------|
| S1 | 14.04 | 2.06 | 0.34 | 4.34 | 0.055 | 58.8 | 6.81 | 1.790 | 5.48 | 4490.0 |
| S2 | 121 | 6.06 | 2.92 | 5.40 | 0.135 | 223.8 | 53.94 | 0.227 | 1.58 | 348.7 |
| S3 | 156 | 6.88 | 3.77 | 6.13 | 0.130 | 277.8 | 47.13 | 0.258 | 1.21 | 303.2 |